\documentclass{article}
\usepackage[utf8]{inputenc}
\usepackage{geometry}
\usepackage{amsmath}
\usepackage{amssymb}
\usepackage{amsfonts}

\title{The Constructivist's Programme and the Problem of Pregeometry}
\date{}

\author{Niels Linnemann\footnote{Institut für Philosophie, Universit\"{a}t Bremen, Postfach 330 440, 28359 Bremen, Germany; niels.linnemann@uni-bremen.de.}\hspace{0.1cm} and Kian Salimkhani\footnote{Philosophisches Seminar, Universit\"{a}t zu K\"{o}ln, Albertus-Magnus-Platz 1, 50923 Cologne, Germany; k.salimkhani@uni-koeln.de.}}

\usepackage[authoryear, round]{natbib}
\usepackage{graphicx}
\usepackage{hyperref}

\usepackage{setspace}


\begin{document}
\sloppy
\maketitle

\begin{abstract}
\noindent
Prominently, Norton (2008) argues against constructivism about spacetime theories, the doctrine that spatiotemporal structure in the dynamics only has derivative status. Among other things, he accuses Brown and Pooley's dynamical approach to special relativity of being merely half-way constructivist: setting up relativistic fields as presupposed in the dynamical approach to special relativity already requires spatiotemporal background structure (pregeometry from now on). We first assess a recent solution proposal by Menon and then provide our very own defense of constructivism along two independent lines.
\end{abstract}

\section{Introduction}
In a series of publications (see \cite{BrownPooley:2001}, \cite{Brown:2005}, and \cite{BrownPooley:2006}), Brown and Pooley proposed a reading of both special relativity (SR) and general relativity (GR) that renders core chronogeometric properties as derivative on fields and their dynamics. According to the dynamical approach to SR, the Minkowski metric is derivative on (i.e., ontologically reduced to) the symmetries of matter field dynamics. According to the dynamical approach to GR, the property of the $g$ field to be surveyed by material rods and clocks (its \textit{chronogeometricity}) is reduced to the coupling behaviour of $g$ to the matter fields, and thus rendered derivative on the dynamics of the interaction between $g$ and the matter fields.\footnote{See, for instance, \citeauthor{BrownReadLehmkuhl} (\citeyear[15]{BrownReadLehmkuhl}). For a different take on the dynamical approach to GR, see \cite{Salimkhani:2020}.}

Prominently, Harvey Brown's and Oliver Pooley's dynamical approach to relativity theory was dubbed `constructivist' by \cite{Norton:2008}: constructivism is the doctrine that spatiotemporal structure is derivative on dynamical laws themselves ``devoid of spatiotemporal presumptions'' \cite[825]{Norton:2008}. Put in more categorial terms, according to constructivism, ``spacetime theories are essentially matter theories'' \cite[821]{Norton:2008}. 

In spite of labelling the dynamical approach `constructivist' by its (alleged) intent, Norton, among other things, accuses Brown of actually offering only a limited form of spacetime constructivism: setting up matter fields, Norton argues, requires some kind of (supposedly already spatiotemporal) background structure in the form of a differentiable manifold, such that \textit{there are} ``components of spacetime geometry that do not supervene on matter and also cannot be explained by matter theories; that is, they would not, in the language of Brown’s text, be a `result of' properties of matter'' \cite[823]{Norton:2008}. Thus, as we shall discuss below, the constructivist is challenged to show that such background structure (`pregeometry' from now on) is either (a) non-fundamental (e.g., derivative on the field dynamics), (b) non-spatiotemporal (the pregeometry does not bring in spatiotemporality), or (c) non-ontic (the pregeometry is not part of the overall ontology at all---i.e., not even of the non-fundamental ontology---but a purely representational artefact). Either of these options would ensure that Norton's concern is moot. Call this challenge on the way to full spacetime constructivism the \textit{problem of pregeometry}.

In this paper, we investigate whether a constructivist take on SR and GR can succeed with respect to the ontological status of the manifold. For the sake of focus, the paper accepts the dynamical approach as a tenable position in other respects.\footnote{See, however, \cite{Norton:2008} for further critical points; generally, influential critical accounts of the dynamical approach include \cite{janssen2002reconsidering}, \cite{Acuna}, and \cite{WeatherallDynamical}.}

We first present Norton's challenge for the constructivist (section 2), and map out possible replies at a general level (section 3). We then assess and criticise the only explicitly proposed solution strategy, namely that by Menon (section 4). Dissatisfied with this response to Norton, we provide two independent lines of defense of spacetime constructivism of our own: one solution strategy is suggested by an account of the dynamical approach due to Stevens (section 5.1), and, the other one arises from attacking Norton's view of the manifold as an indispensable object (section 5.2). Here, we also comment on often ignored statements by Brown himself.

\section{The Problem of Pregeometry}
In this section, we introduce the `problem of pregeometry' for the constructivist in more detail. Take the constructivist to claim---\textit{pace} the standard fundamentalist\footnote{\cite{Norton:2008} takes the SR constructivist to deny what he dubs the `realist conception' of SR spacetime: 
\begin{quote}
\begin{enumerate}
    \item[(1)] There is a four-dimensional spacetime that can be coordinatized by a set of standard coordinates $(x, y, z, t)$, related by the Lorentz transformation. 
    \item[(2)] The spatiotemporal interval $s$ between events $(x, y, z, t)$ and $(X, Y, Z, T)$ along a straight line connecting them is a property of the spacetime, independent of the matter it contains, and is given by $s^2 = (t-T)^2 - (x-X)^2 - (y-Y)^2 - (z-Z)^2$ [\dots]
    \item[(3)] Material clocks and rods measure these times and distances \textit{because} the laws of the matter theories that govern them are adapted to the independent geometry of this spacetime. \cite[823]{Norton:2008}
\end{enumerate} 
\end{quote}
However, \cite{Menon} (and also \cite{North:2017}) propose that all the proponent of spacetime constructivism \textit{actually} has to maintain is that spacetime is \textit{non-fundamental}. Note, however, that denying the reality of (parts of) (1)--(3) also remains a viable option for a constructivist. Indeed, we shall ultimately defend such a view in section~4.} conception of spacetime---that \textit{all} spatiotemporal properties are \textit{fully} reducible to field dynamics; in particular, that even topological properties are fully reducible to field dynamics. The problem(s) of pregeometry then take the following forms:

\begin{description}
\item[Special-relativistic problem of pregeometry] The dynamical approach ontologically reduces the Minkowski metric to matter field dynamics. The dynamical approach fails to reduce core spatiotemporal properties that are not specifically associated to metric structure but which are required to set up the matter fields in the first place.\footnote{This is to say that the mathematical representation of physical fields requires recourse to a manifold background structure: they are defined as maps from a manifold onto respective target spaces.} These core spatiotemporal properties are expressed in the assumption of a \textit{joint} manifold structure, called pregeometry.
\item[General-relativistic problem of pregeometry] The dynamical approach ontologically reduces the chronogeometric property of the $g$ field---a core spatiotemporal property---to the dynamics of the $g$ field and the matter fields (including their mutual interactions).\footnote{Let us add to the standard presentation of the dynamical approach proponent that the dynamical approach arguably also reduces the $g$ field's \textit{causal} property---which is a core spatiotemporal property ontologically prior to chronogeometricity (if $g$ has chronogeometricity, it also has the causal property)---to the mutual interactions of the $g$ field and the matter fields. The causal property of $g$ is its property that matter fields and test particles can only move in accordance with its conformal structure.} The dynamical approach fails to reduce core spatiotemporal properties which are required to set up the fields in the first place. These core spatiotemporal properties are expressed in the assumption of a joint manifold structure, called pregeomety.
\end{description}
\noindent
As a consequence, the dynamical approach can\textit{not} be considered full-fledged constructivist---call it \textit{half-way constructivist}. Take the case of SR: Brown and Pooley---not pursuing the project of reducing all spatiotemporal structure but merely metric structure to the dynamical properties of the matter fields---admit\footnote{See \cite[829]{Norton:2008} and \cite{Pooley:2013}, respectively.} that they do presume a pregeometric differentiable manifold structure. To count as full-fledged constructivist, the dynamical approach would have to render \textit{all} (putatively) spatiotemporal structure---not only metric structure---derivative on the dynamics. As the metric structure in SR already is derivative on the dynamics, this essentially amounts to either rendering manifold structure derivative on the dynamics as well, or establishing why manifold structure should not count as spatiotemporal in the first place---neither of which can succeed, \cite{Norton:2008} argues: ``[t]he construction project must tacitly assume an already existing spacetime endowed with topological properties, so that it can introduce spatiotemporal coincidences, and a unique set of standard coordinates $(x, y, z, t)$'' \cite[824]{Norton:2008}. Here is a way to understand Norton's worry: suppose for instance that all we have are two matter fields $\phi_1$ and $\phi_2$ each defined on manifolds $M_1$ and $M_2$ respectively. We still expect our theory to answer a question of the following form: `The value of field $\phi_1$ at a point $P_1$ in the manifold $M_1$ is $\phi_1(P_1)$, what is the value of field $\phi_2$ at this same point?' To be able to answer such a question, we need to employ some further assumption, the obvious choice being that $M_1$ and $M_2$ are identical, such that $P_1$ and $P_2$ are identical (see also Fig.~\ref{figure:pincushion}).\footnote{It is worth pointing out that, recently, \cite{Binkoski} has identified this issue of `geometric coordination' as a problem for any programme that attempts to render spacetime as a structural quality of each field separately---and so not necessarily only of the dynamical approach.}

Now, Norton takes it that what the dynamical approach needs to assume is a manifold structure---the pregeometry---that is already spatiotemporal and part of the fundamental ontology. So, if Norton is right and the dynamical approach does have to assume a fundamental manifold, one might ask: what is wrong with this half-way constructivism? Although such a version of constructivism may seem `unfinished' and unsatisfactory\footnote{At least, if some form of relationalism is a driving force behind constructivism because the remaining structure is most straightforwardly interpreted as substantival. Note that relationalism is not what \cite{Brown:2005} is after.} one might argue that this is just how the world is. So why worry? Indeed, the constructivist may actually accept Norton's charge that some spatiotemporal structure is presumed, but argue that this `half-way' approach is a novel and interesting position nonetheless, especially regarding its ontology; in particular, the ontological commitments of half-way constructivism do differ from the standard fundamentalist conception.\footnote{They may, for example, be cashed out in terms of some form of \textit{spacetime emergentism}; see \cite{Martens:Emergent}.} This essentially amounts to accepting Brown's dynamical view as it stands. We shall briefly come back to this option of biting the bullet at the end of this paper. It is important to point out, however, that we do take Norton's claim to be controversial. So the main project of this paper is to show that the dynamical approach is \textit{not} committed to accept the manifold as ontological at all.

\section{Replies to the Problem of Pregeometry}
Recall that Norton claims that the dynamical approach needs to presume the manifold as a spatiotemporal part of the fundamental ontology. So the constructivist is challenged to show that the manifold is, in fact, either (a) non-fundamental (e.g., derivative on the field dynamics), (b) non-spatiotemporal (it does not bring in spatiotemporality), or (c) non-ontic (it is not part of the overall ontology at all, but a purely representational artefact or `scaffolding structure' to start the constructivist project). Either of these options would adequately respond to the challenge.

In this paper, we will consider the following strategies explicitly: (1) solving the problem of pregeometry by explicitly showing that the manifold is, in fact, derivative on the dynamics as well; (2) arguing why manifold should not count as spatiotemporal; (3) based on the distinction (a) to (c), arguing that the manifold is just auxiliary structure.

Route 1 can be read as the attempt to establish that the manifold should never have been counted as `pre'geometry in the first place, because it is just as well derivative on the dynamics, i.e., the manifold is argued to be ontologically reducible. This strategy is discussed in section \ref{sec:Menon}.

Route 2 seeks to demonstrate that the manifold does not bring in spatiotemporality. This arguably requires a clear conception of non-spatiotemporality as opposed to spatiotemporality. We shall discuss this in more detail in section \ref{sec:Stevens}.

Route 3, on the other hand, argues that the manifold is, in fact, just non-ontic auxiliary structure, either by (a) explicitly showing that the manifold is dispensable (we really only use it because it is convenient), and, thus, ontologically irrelevant, or by (b) showing that the manifold is---albeit \textit{in}dispensable in the formal set-up---\textit{dispensable} from the actual physical set-up (in other words, it does not appear in the ontology). Notably, this route is independent of whether the manifold is ontologically reducible or not.

Note that another form of reply is arguably to simply deny that there is any issue, that is to simply accept half-way constructivism. We do acknowledge that half-way constructivism is a tenable option that has the advantage of being independent of additional philosophical assumptions regarding spatiotemporality or ontological status. When adopting half-way constructivism, one does not see Norton's pregeometry challenge as threatening the dynamical approach as such, but as merely demanding the dynamical approach to make its stance on the manifold clear. 

\section{Debunking Menon's Algebraic Approach}
\label{sec:Menon}
For his (dis)solution of the problem of pregeometry, \cite{Menon} makes use of the algebraic reformulation of GR going back to \cite{Geroch1972}. Whereas fields including the $g$ field are standardly represented as structures on a manifold, and thereby presuppose a pregeometric structure, the basic idea here is to use the rich algebraic structure of the fields (that is how they can be combined, i.e., multiplied, added, etc.) to, in the end, do away with the manifold structure at the fundamental level and make it derivative on the field dynamics.

For reasons of simplicity, we will---similarly to Menon---only consider scalar fields in the following.\footnote{A more general account builds on that of Einstein algebras \`{a} la \cite{Geroch1972}. See \cite{BainHoleArgument} for a philosophical introduction.} Then, the set of infinitely differentiable real-valued fields $C^{\infty} (M)$, usually understood as a set of fields on a manifold, can be reconceptualised as the abstract algebra underlying it. The manifold version is then just a specific representation of this algebra; notably, using less than this (abstract) algebraic structure does not suffice (see \citeauthor{Menon} (\citeyear[1277]{Menon})).

Now, the interpretational move of Menon is as follows: He takes the (abstract equivalent of a) field to be fundamental on the algebraic view (as an element of an algebra, and thus as an abstract entity). In contrast, the manifold points are reconstructed from these abstract entities and thereby rendered \textit{derivative}. If this holds, one could say that the (then only so-called) pregeometry---the manifold---is actually ontologically reducible to a non-spatiotemporal algebraic structure in terms of which the dynamics are expressed fundamentally.

Using the well-known theoretical equivalence between the manifold and the algebraic representation,\footnote{See \cite{Ryziencska} and \cite{Rosenstock:2015dka}.} we now argue that either the `how much pregeometry needed' question (1) only sets up a pseudo-problem (it is only not a problem in the algebraic representation because it is no problem in \textit{any} representation, i.e., it is not a problem at all), or (2) poses a genuine problem that cannot be transformed away through a mere change of representation. We can indeed see quite explicitly that all that Menon aims to do in the algebraic representation can already be done---if at all---in the standard representation (the manifold representation), as we show now. 

Menon distinguishes between kinematically possible models (KPMs) of two kinds at the level of manifold structure:
\begin{quotation}
    \noindent
    KPMs of relativistic theories are tuples of the form $\langle M, g_{\mu\nu}, \phi_i \rangle$, where $M$ is a smooth manifold, $g_{\mu\nu}$ is a Lorentzian metric tensor, and $\phi_i$ is a placeholder for matter fields.~\dots

    Models of this kind implicitly \dots\ have an extra bit of structure---a link between the manifold and the fields. Specifically, for any point in $M$, distinct fields, $\phi_1, \dots, \phi_n$, can be `evaluated' at that point, and their values can then be taken \dots\ to represent properties of the same space-time point. Call a KPM in which this link does exist a \textit{KPM of the first kind}.
    
    \dots
    
    Given the arbitrariness of the coordinate function, there is no \textit{a priori} need for the coordinatisation of two composite functions to coincide. In other words, there is no \textit{a priori} reason to require that the $x$ in $\phi(x)$ is the same as the $x$ in $\psi(x)$---indeed, our use of the same variable to denote both is indicative of a failure to allow for this mismatch. Call models in which preimages of $x$ in $\phi(x)$ and $\psi(x)$ can represent different points, \textit{KPMs of the second kind}. \cite[1274--1275]{Menon}
\end{quotation}
So, KPMs of the first kind presuppose a joint reference structure (`fields living on one and the same pregeometry') whereas KPMs of the second kind, at least \textit{prima facie}, do not (`fields possibly living on completely independent pregeometries'). 

KPMs of the first kind are naturally linked to a manifold representation (see quote). Now, Menon's intuition is that KPMs of the second kind are \textit{not} linked to a manifold representation. On this picture, showing KPMs of the first kind to be derivative on KPMs of the second kind would then show that the manifold is derivative, too. (As Menon points out ``[t]he central aim of this paper is to demonstrate that KPMs of the second kind can be constructed for SR'' \cite[1275]{Menon}, and that ``Norton’s criticism is based on assuming the KPMs at play can only be of the first kind'' \cite[1275]{Menon}.) But we have just seen that even Menon himself shows that KPMs of the second kind \textit{can}, after all, be constructed in the manifold representation: we just need to accept that we then have multiple distinct manifolds, i.e., one for each field, in our KPMs. So, even if Menon's story about how to obtain the KPMs of the first kind from the KPMs of the second kind in the algebraic formulation is correct, an exact counterpart can be told in the manifold representation. Without any additional argument, it is not clear how mere re-representation solves the ontological issue at stake. (In fact, Norton's very criticism is \textit{based} on considering the option of there being KPMs of the second kind in the manifold representation in the first place: arguing that we can accept several distinct manifolds for SR does not tell against Norton's key argument that we are forced to accept a single manifold structure as the result of a common origin inference.)

Some explication is in order as to why mere equivalence between manifold and algebraic representation is not enough: both manifold and algebraic representation seem to go along with their own (natural or formalistic) interpretation but some external reason is needed as to why one is preferred over the other---and, arguably, even to demonstrate that there should be two distinct as opposed to one joint interpretation of the two equivalent representational structures to begin with. Now, the dynamical approach proponent is generally committed to denying that a formal equivalence between statements is tantamount to that there is a joint interpretation.\footnote{Recall the debate between, on the one hand, the different priority readings by, for example, \cite{BrownPooley:2001} (as proponents of the dynamical approach) and \cite{BalashovJanssen:2003} (as proponents of the geometrical approach), and, on the other hand, a joint reading according to \cite{Acuna}.} After all, the basic idea behind the dynamical approach to, say, SR is to think of the metric as derivative on the matter field equations defined on the manifold, \textit{although}, purely mathematically speaking, stating that equations of motions (written in coordinates) display Lorentz-invariance is equivalent to that the relevant background structure is a Minkowski metric. The difference for a dynamical approach proponent, however, between being entitled to asymmetrically interpret the equivalence just sketched, and not being so for the formal equivalence between algebraic and manifold representation derives from that in the first case one can adhere to a basic posit of the dynamical approach (`explanations should be dynamical not geometrical') whereas in the second case a corresponding posit does not offer itself immediately, at least not from the dynamical approach perspective specifically. At the same time, simply saying that the algebraic take is preferred over the manifold one because the former is of non-spatiotemporal nature, is a \textit{petitio principii}; the dynamical approach proponent must show why her view leads to full constructivism, not argue that it is compatible with it.\footnote{We have recently become aware of a criticism of \cite{Menon} by \cite{ChenFritz} to the effect that his approach privileges scalar fields in terms of which all other kind of fields have to be expressed. (This criticism is arguably based on an uncharitable reading of Menon who could very well be taken to only focus on the scalar model as a toy model; in this sense, we view \cite{ChenFritz} as a genuine continuation of Menon's proposal, not a rectification of it.) Superficially, this seems to complement our sentiment that the algebraic structure is just the manifold in disguise: their observation after all entails that the algebraic structure linked to scalar fields is some background structure for (the other) fields. However, we insist that a mere re-representability of other fields purely algebraically---as \cite{ChenFritz} aspire to do---suffers from the same issue just described for Menon's scalar model.}

In any case, we do acknowledge, however, that Menon (indirectly) gives an account of under which circumstances we can assume coincidence of coordinates, that is, in other words, assume a unified ordering structure for all fields. His explanation (already mentioned above) goes as follows:
\begin{quote}
    Consider, now, a coordinate system on which we have good reason to believe that, say, a Gaussian wave packet of the $\phi(x)$ field bounced off a Gaussian wave packet of the $\psi(x)$ field in the neighborhood of some point. There will be a class of coordinatizations [i.e., functions from $\mathcal{M}$ to ${\rm I\!R}^n$; our addition] that assign field values in such a way that the dynamics of each field determines that a collision took place in the vicinity of some point and another class of coordinatizations on which the kinks in the trajectories (or, more generally, some fact about the dynamical interaction) of each particle do not take place at the same coordinate value. For various reasons, the former class of coordinate systems might be preferable. For free fields, there simply is no operational sense of spatiotemporal point coincidence, although such points can still be defined. In such a case, though, these would amount to arbitrary stipulations. \cite[1279]{Menon}
\end{quote}
This explanatory account for point coincidences can only be adhered to if one has already accepted a scenario of a multitude of manifolds---one for each field. But then again it seems as if we have to either dig deeper to find out how these various manifolds are derived from the dynamics (and, thus, even if spatiotemporal, are only derivatively so)---or come up with some argument as to why the manifolds are non-spatiotemporal. In particular, it seems natural to ascribe the same spatiotemporal status to all manifolds that are assumed: either all manifolds are equally spatiotemporal, or none of them are. Notably, the former option of assuming several spatiotemporal manifolds seems disadvantageous to assuming a single spatiotemporal manifold. 

The upshot then is the following: we may accept that our ability to assume multiple \textit{non-spatiotemporal} manifolds is a solution to the problem of pregeometry\footnote{Again, the task was to do without an assumed spatiotemporal manifold structure.} \textit{because} the commitment to multiple \textit{spatiotemporal} manifolds is implausible (whereas it is supposedly plausible to have multiple \textit{non-spatiotemporal} manifolds). However, Norton can still maintain that the commitment to a single (spatiotemporal) manifold is more plausible via insisting that manifold structure \textit{is} spatiotemporal: if manifold structure is spatiotemporal, it is better to assume a single spatiotemporal manifold than a plethora of those (which, again, is deemed implausible). Hence, one is back to the issue we discussed in the previous section: whether manifold structure \textit{as such} is spatiotemporal (regardless of how many manifolds are considered); here, one will be able to draw on our independent proposal from the next section that manifold structure lacks essential features of the spatiotemporal---in particular, a difference between space and time.

In any case, to thoroughly address Norton's counter argument, the constructivist arguably has to additionally show that multiple non-spatiotemporal manifolds would even merely be representational. For, while Occam's razor does not exclude having several non-spatiotemporal and merely representational ordering structures, it might exclude having several non-spatiotemporal and ontologically relevant ordering structures, and does certainly exclude having several spatiotemporal and ontologically relevant ordering structures---especially, when they exhibit point coincidences that allow for a common origin inference (see \cite{Norton:2008}; see also \cite{janssen2002reconsidering} and \cite{BalashovJanssen:2003}).\footnote{At first sight, there being several manifolds squares well with the constructivist claim that a manifold is mere representational, non-spatiotemporal ordering structure. On the other hand, the fundamentalist claim that manifold structure is fundamental and spatiotemporal will receive further support, if it is demonstrated that this claim is most plausible for a \textit{single} manifold; this is exactly what Norton achieves with his common origin inference, aimed against the constructivist. If manifold structure was mere ordering structure, he argues, it would seem odd that the project is most successful when a \textit{common} ordering structure is assumed. The success or plausibility of full constructivism should not depend on a specific choice for the ordering structure.} We come back to this below.

\section{Two new solutions to the problem of pregeometry}

\subsection{The problem of pregeometry in light of Stevens' Regularity Relationalism}
\label{sec:Stevens}
In this section, we argue that the conceptualisation of the dynamical approach by \cite{Stevens:2018} inspires a reply to the problem of pregeometry that is mostly in line with route 2. It must be clearly stressed that at no point Stevens himself makes reference to the pregeometry problem; in fact, he himself explicitly sees what we call pregeometry as `spatiotemporal sub-metrical structure' \cite[357]{Stevens:2018} and so apparently has not---at least at time of writing---taken up a non-spatiotemporal reading of the manifold as non-spatiotemporal which we take to be heavily suggested, or so we will argue, by his own account.

As presented above, the general task for the constructivist is to pull spacetime out of the dynamics governing the fields without already presuming spatiotemporal structure. For this, we can learn from Stevens and his analogy between the role of the Leibniz group in Newtonian mechanics and the role of the manifold structure in the dynamical approach. The analogy draws on the fact that in both contexts the fundamental setting has less structure than the actual dynamical setting of the theory: in the case of a Leibnizian relationalism on Newtonian mechanics, there are full Leibnizian symmetries at the fundamental ontological level, while at the level of the actual dynamics only Galilean symmetries reside. Thus, the less symmetric dynamical structure is supposed to supervene on the more symmetric fundamental structure (more symmetry means less structure).\footnote{Notably, this move has be seen as rather unsatisfying or even as a dubious sleight of hand (e.g., \cite{Belot:2000}).} In the case of the dynamical approach to SR, there are matter fields on a manifold \textit{without metric} at the fundamental ontological level, while at the actual dynamical level all matter fields move in a Minkowski spacetime, i.e., a manifold \textit{with metric}.

More concretely, Leibnizian spacetime---unlike Galilean spacetime---is invariant under linear accelerations, i.e., has no affine structure. As \cite{Pooley:2013} puts it, in order to accommodate for more structured (less symmetric) dynamics, the Leibnizian relationalist could either (1) extend her ideology, that is the set of relations between the fundamental entities of the theory, or (2) move to a different theory committed to different fundamental entities, or (3) tell a story on how to ``have one's cake and eat it'' \cite[564]{Pooley:2013}. A well-known example for option (3) is Huggett's (\citeyear{Huggett:2006}) \textit{regularity relationalism}, where the dynamical laws of Newtonian Mechanics described in Galilean spacetime supervene on a Humean mosaic in Leibnizian spacetime;\footnote{Since Leibnizian spacetime is fundamental, Huggett's approach is not a fully constructivist project either.} not only the actual dynamics but \textit{also} that it is described in Galilean spacetime arises as a liberalised Humean best-systems analysis (later known as `super-Humeanism'\footnote{See for instance \cite{Lazarovici}, and Pooley's (\citeyear{Pooley:2013}) `have-it-all relationalism'.}). Due to the analogy between option (3) and the general programme of constructivism, the lessons of Huggett's account can be put to use in the pregeometry debate when trying to have spacetime supervene on non-spatiotemporal structure (i.e., route 2).

Now, \cite{Stevens:2018}, with Huggett's proposal for Leibnizian relationalism at hand, cashes out the aforementioned analogy between Leibnizian relationalism and the dynamical approach by arguing that the pregeometric structure is simply that structure with respect to which the patterns of a Humean mosaic are formulated. Concretely, a successful constructivist project would take a material world with the pregeometric structure as the supervenience basis and then best-systematise the dynamical laws in terms of a (more) spatiotemporal structure that supervenes on it \cite[]{Stevens:2018}.

At first sight, this supervenience basis might be conceived of as a fundamental \textit{spatiotemporal} supervenience basis. As a matter of fact, Stevens himself explicitly speaks of `spatiotemporal sub-metrical structure' \cite[357]{Stevens:2018}.

Our point now is simple but forceful: rendering pregeometry as the complete supervenience basis \textit{for everything else} is tantamount to making the pregeometric structure independent of any field that might be added (or, put technically, defined) on top of it. So on this account the pregeometric structure is a fundamental, i.e., ontologically independent entity.\footnote{We take it to be analytic to the Humean/regularity relationist understanding of supervenience basis that what is conceived as supervenience basis is ontologically independent of the objects and patterns that are supposed to be described through it. Of course, that might be questioned but then the whole account of Stevens looks questionable to begin with.} But once this is realised, the question immediately becomes as to how one could understand a mere manifold structure as spatiotemporal. 
First, it is essential to space and time (or spacetime) to play the role of an ordering structure. If the different concepts of time from physics to psychology and phenomenology have anything in common, it is the idea that time, among other things, is an ordering parameter. The same holds for space. Second, it is essential to space and time that the two are in a relevant sense distinct from one another.\footnote{Note that the commitment to essentialism here is relatively modest: one only commits to a minimal set of properties anyone with essentialist intuitions towards space and time would want to agree on (or at least, there do not seem to be relevant positions that are thereby excluded).}
Given this view on the spatiotemporal, the manifold cannot count as spatiotemporal: \textit{it is an ordering structure but lacks a distinction between one ordering parameter as opposed to the others}.

\subsection{The Problem of Pregeometry as a Problem of Indispensability}
\label{pre:indisp}
In the previous sections we argued that Menon's algebraic attempt at solving the problem of pregeometry---as such, an instance of route~1: establishing manifold structure as derivative on the algebraically formulated dynamics, and, thus, establishing manifold structure as fundamentally dispensable---is ill-directed, and that Stevens's conceptualisation programme of the dynamical approach can be read to inspire a solution that depends on a concrete notion of the spatiotemporal---similar to route~2: showing that manifold structure is, albeit indispensable (and arguably also fundamental), non-spatiotemporal. In the following, we shall dwell further on how to combine these insights to be able to counter Norton's central line of argument.

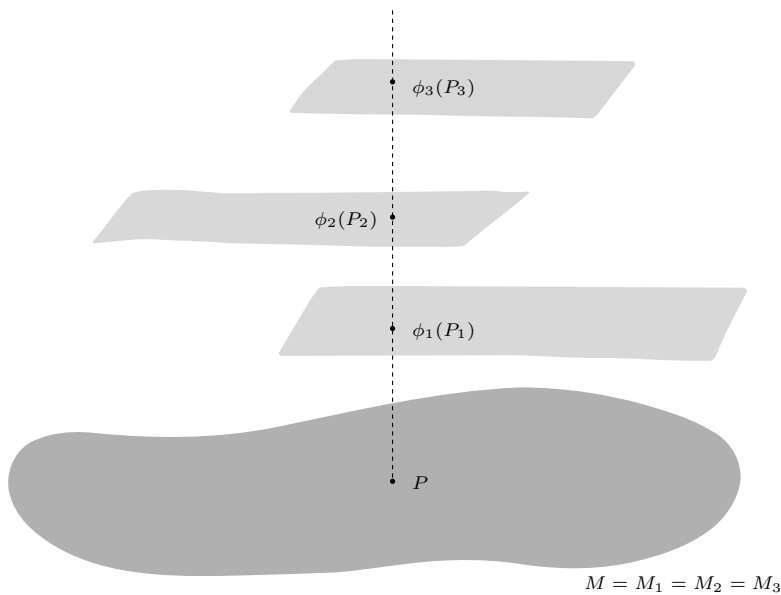
\begin{figure}[ht]
	\begin{center}
	    \vspace{7pt}
		\includegraphics[width=0.65\textwidth]{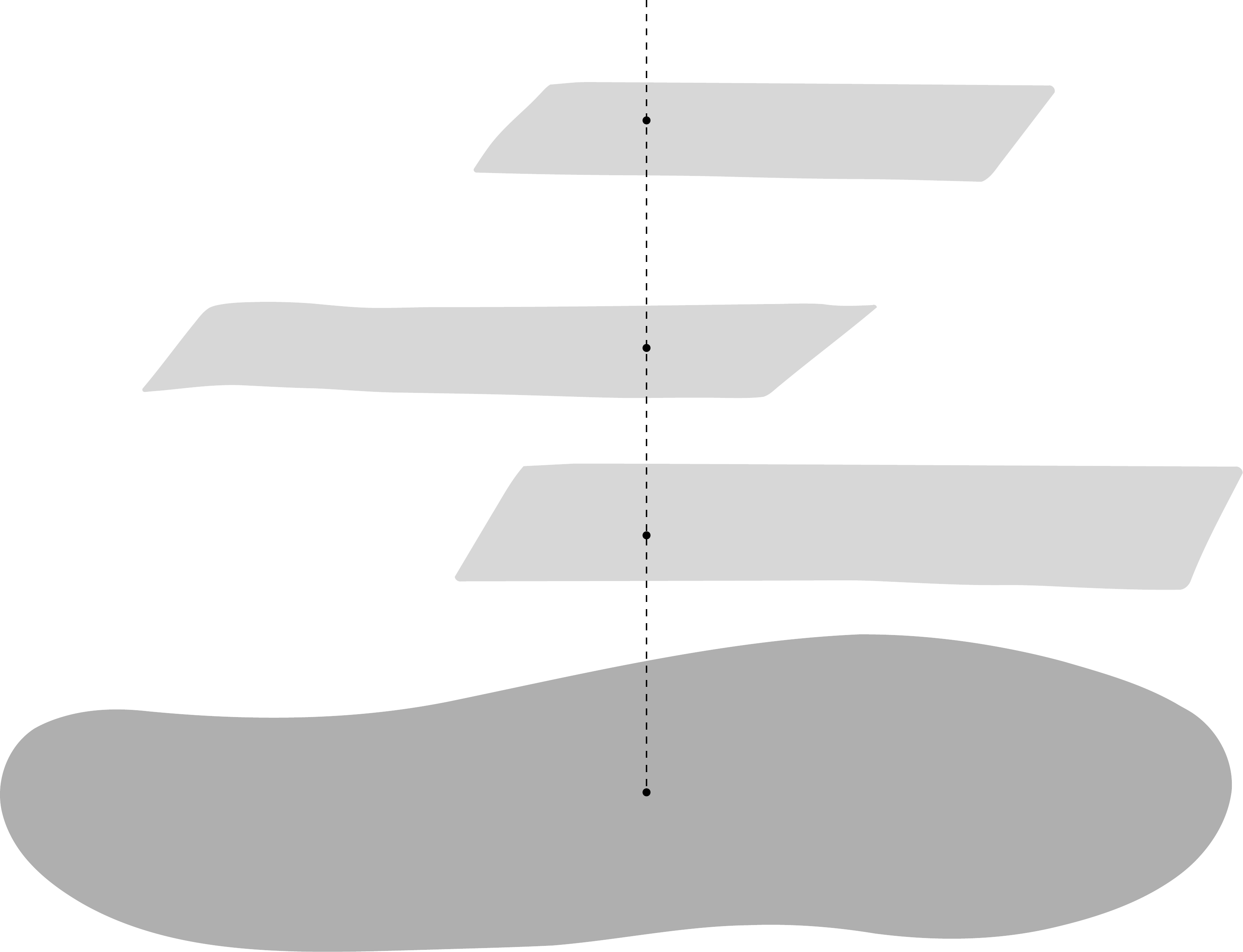}
		\vspace{7pt}
		\put(-65,-5){\scriptsize $M = M_1 = M_2 = M_3$}
		\put(-130,33){\scriptsize $P$}
		\put(-130,91){\scriptsize $\phi_1(P_1)$}
		\put(-167,133){\scriptsize $\phi_2(P_2)$}
		\put(-130,183){\scriptsize $\phi_3(P_3)$}
		\caption[Visualisation of Norton's indispensability argument for the manifold]{\label{figure:pincushion} Visualisation of Norton's indispensability argument for the manifold $M$ (dubbed here the `pincushion rationale'). Given some matter fields $\phi_i$ defined on individual manifolds $M_1$, $M_2$, $M_3$, respectively, only identifying all individual manifolds with the single manifold $M$ (the `pincushion') allows to compare field values of different fields at a point $P$.}
	\end{center}
\end{figure}

In particular, recall that Menon's account entails that fields, at first defined on distinct manifolds, can be taken to refer to one and the same manifold through considering their dynamical evolution; after all, the different fields are not free, but interact. This result addresses Norton's \textit{coincidence concern}, namely that ``[t]he construction project must tacitly assume an already existing spacetime endowed with topological properties, so that it can introduce spatiotemporal coincidences, and a unique set of standard coordinates $(x, y, z, t)$'' \cite[824]{Norton:2008}. Figuratively speaking, one can dub Norton's position the `pincushion rationale' (see Fig.~\ref{figure:pincushion}): evaluating the field values of different fields at a point involves fixing these fields to one and the same background structure just like a needle fixes different layers of fabric to a pincushion. As a consequence, the rationale claims that one is committed to (what serves as) the `pincushion' as part of the fundamental ontology---this is Norton's indispensability argument for the manifold.

\noindent 
However, committing to the `pincushion' might simply yield too much: it is the \textit{needles} (i.e., the field dynamics) that fix the coincidences, not the pincushion which additionally fixes an `absolute' localisation in terms of associating a spacetime point to each `needle'. When all we need is coincidences, as Norton argues himself, why should we be forced to commit to the `pincushion'? Thus put, another way to block Norton's coincidence concern over and above that proposed by Menon is---or, is at least motivated by---what we call the `needles only rationale'. Speaking less figuratively, what ultimately is at stake is whether or not the coincidence concern translates to an indispensability argument for the \textit{manifold of spacetime points} specifically, and whether the manifold's alleged ontological status as well as its being spatiotemporal follow from it.

Notably, \cite{Brown:1997} similarly remarks that

\begin{singlespace*}
\begin{quote}
[i]t is tempting to assign such autonomy [`reality' in \cite[156]{Brown:2005} or, in our terminology, `ontological independence'; our remark] to the continuum of space-time points precisely because a dynamical field is standardly defined as a map from this continuum into a suitable set of tensors of a certain kind. The temptation is closely related to the intuition which led Lorentz to maintain the notion \dots\ of an imponderable \dots\ ether, whose ultimate role was to provide ``a peg to hang all these things [electromagnetic fields] on''. Lorentz's ether failed in the minds of most physicists to survive the developments of 1905, and in particular Einstein's persuasive demonstration of the ``redundancy'' of the electromagnetic medium, ponderable or otherwise. Yet arguably the core of Lorentz's intuition was still widely, if unwittingly, adhered to: the ``peg'' \dots\ now became the space-time manifold itself. 
\cite[68]{Brown:1997}
\end{quote}
\end{singlespace*}
Brown then goes on and stresses, once again, that the debates in the context of diffeomorphism invariance and the hole argument in general relativity call ``into question the reality of the post-Lorentzian peg. In pre-quantum physics then, space-time points are perhaps best viewed not as entities in their own right, but as correlations or links between the individual degrees of freedom of distinct physical fields'' \cite[68]{Brown:1997}. Arguably, this is a variant of our `needles only rationale' which, in this case, is bolstered by the observation that we usually do not commit to gauge redundancies as physically real (see also \citeauthor{Brown:2005} (\citeyear[156]{Brown:2005})). We take it that this marks an important sense in which the coincidence concern does \textit{not} translate to an indispensability argument for the manifold specifically (see also below).

Now, the `needles only rationale' seems to be easily countered by the insight that the `totality of needles', or the totality of what Brown calls `correlations' or `links' \cite[68]{Brown:1997}, may just reflect the manifold structure in disguise, so that nothing really is gained by this argument. However, or so we insist, it reflects the manifold structure only in so far as that the manifold provides some representational structure for the \textit{physical} events:
manifestly, the `totality of needles' is the totality of (interaction) events, \textit{not} the totality of spacetime points; it is a manifold of `needles' or correlations, not a manifold of spacetime points (the latter of which, again, would additionally fix an `absolute' structure that is \textit{only afterwards} argued to bear a certain redundancy in virtue of diffeomorphism invariance and the hole argument).\footnote{One way of making the notion of `totality of interaction events' formally explicit is presented by \cite{WestmanSonege} and taken up by \cite[\S9.4]{Hardy}: The basic idea on the Westman-Sonego approach is to form a space of local beables spanned by values corresponding to scalar contractions of the fields; each point of the space is a physical event.}

Each needle's referring to a manifold point is merely an (arguably natural) \textit{ex post} interpretation---manifestly, it still just marks the correlation between different field values. For instance, a supernova event gives physical correlations between the gravitational, the electromagnetic, and other fields (like neutrino fields) that can be used to fix their relative displacement in a certain neighbourhood. The fact that we can associate such correlations of gravitational, electromagnetic and other physical events with a particular point of the manifold of spacetime points is not a fact of what the world is like, but a representational result: the manifold of spacetime points is (just) a convenient way to organise this information.

In fact, note that for this to work we are not even committed to a \textit{manifold} of needles: not every event needs to be an interaction of different matter fields and correspond to a point on the manifold, we just need an adequate (possibly just discrete) network of interaction events such that we can coordinate the different field structures.\footnote{In fact, one may spell out the `needles only' perspective for GR further by appeal to results on the causal structure in GR from \cite{Malament1977} and, more concretely, a potential successor theory of GR based on these results, namely causal set theory (see \cite{Wuthrich2012} for an introduction): causal set theory posits a set of elementary events (in some sense corresponding to the `needles'), which combined appropriately yield general-relativistic spacetime.} 

As a result---\textit{pace} Norton---the manifold is, indeed, \textit{dispensable} from the ontology, i.e., non-ontic. Notably, independently of whether the manifold is dispensable from the formalism or shown to be derivative on something else.

\section{Conclusion}
Let us briefly retrace our steps. We agree with Norton that the problem of pregeometry for the dynamical approach has generally received too little attention, despite its centrality. We have therefore appraised the only existing solution and proposed two new ones. 

We started out by evaluating Menon's proposal: his account is unfortunately not convincing without further justification as to why one of two equivalent representations is ontologically more relevant.\footnote{Arguably, this issue may be remedied by considering successor theories, on which also \cite{Menon} ultimately seems to rely (see \cite{HuggettLizziMenon:2020}).} 

We then worked out two independent proposals of our own. The first account is inspired by Stevens' conceptualisation of the dynamical approach as regularity relationalism: a best-systems account like that of regularity relationalism implies that the supervenience basis is independent of what supervenes; on such a view of the dynamical approach, the manifold---playing the role of the supervenience basis---must then be seen as ontologically independent of its field content. Whether or not the manifold is spatiotemporal, must be independently decided for the manifold. Given now that spatiotemporality minimally involves a distinction between space and time, the manifold structure---the pregeometry---can only be read as \textit{non}-spatiotemporal. The second account builds on the insight that what Norton essentially puts forward is an indispensability argument for the manifold (notably, on the basis of physical, not purely formal considerations). We countered that the manifold's indispensability is, in fact, \textit{not} straightforwardly obtained: Norton's central argument only establishes the indispensability of point coincidences, which Norton then takes to imply the indispensability of the manifold. Against this, we argued that point coincidences do, in fact, suffice, if modelled in terms of interactions \`{a} la Menon (which is notably independent of his algebraic approach). 

Whether our first or second reply is to be favoured is, arguably, a question of background assumptions. We favour the dispensability account: after all, the dispensability account of the manifold is exactly in line with how practitioners of quantum gravity think dismissively of the manifold. Conversely, when seen from the vantage point of the practitioner, the problem behind the regularity relationalist account is a too literal view of the manifold. Now, discrepancy with practice need not be a tie-breaker in a metaphysical consideration; in any case, siding with practitioners means taking issue with a regularity relationalist account of pregeometry and thus effectively with a conceptualisation of the dynamical approach in terms of regularity relationalism à la Stevens.

\section*{Acknowledgements}
We thank Andreas Bartels, Stefan Heidl, Nick Huggett, Andreas Hüttemann, Rasmus Jaksland, Dennis Lehmkuhl, Niels Martens, Matthew Parker, Christopher Smeenk, and Christian Wüthrich for helpful discussions and comments on earlier versions of this paper. We also thank the audiences of the \textit{Chicago-Geneva Beyond Spacetime seminar} (May 2018), the \textit{7th Biennial Conference of the European Philosophy of Science Association} in Geneva in September 2019, the \textit{5th Annual Conference of the Society for the Metaphysics of Science} in Toronto in November 2019, and the Rotman reading group in March 2021. The work of K.S.\ and N.L.\ was supported by the German Research Foundation (DFG, grant number: FOR 2495). In addition, the work of K.S.\ was performed under a collaborative agreement between the University of Illinois at Chicago and the University of Geneva, and was made possible by grant number 56314 from the John Templeton Foundation; and the work of N.L. was supported by the Swiss National Science Foundation (105212\_165702). K.S.\ is very grateful for support of the DFG research unit \textit{Inductive Metaphysics}, and for a Beyond Spacetime Junior Visiting Fellowship awarded by the Geneva symmetry group. N.L.\ is very thankful for a research fellowship awarded by the DFG research unit \textit{Inductive Metaphysics}, and the mentioned funding by the SNF.

\bibliography{references}
\end{document}